\input{aipcheck}

\documentclass[
    ,final            % use final for the camera ready runs
  ]
  {aipproc}

\layoutstyle{6x9}

\begin{document}

\title{Shareholding Networks in Japan}


\author{Wataru Souma}{
  address={ATR Network Informatics Laboratories, Kyoto 619-0288, Japan}
}

\author{Yoshi Fujiwara}{
  address={ATR Network Informatics Laboratories, Kyoto 619-0288, Japan}
}

\author{Hideaki Aoyama}{
  address={Department of Physics, Graduate School of Science, Kyoto University, Kyoto 606-8501, Japan}
}

\begin{abstract}
The Japanese shareholding network existing at the end of March 2002 is studied empirically.
The network is constructed from 2,303 listed companies and 53 non-listed financial
institutions. We consider this network as a directed graph by drawing edges from
shareholders to stock corporations. The lengths of the shareholder lists vary
with the companies, and the most comprehensive lists contain the top 30 shareholders.
Consequently, the distribution of incoming edges has an upper bound, while that of outgoing edges
has no bound. The distribution of outgoing degrees is well explained by the power
law function with an exponential tail. The exponent in the power law range is $\gamma=1.7$.
To understand these features from the viewpoint of a company's growth,
we consider the correlations between the outgoing degree and the company's age, profit, and
total assets.
\end{abstract}

\maketitle

%%%%%%%%%%%%%%%%%%%%%%%%%%%%%%%%%%%%%%%%%%%%
%% MAINMATTER
%%%%%%%%%%%%%%%%%%%%%%%%%%%%%%%%%%%%%%%%%%%%

\section{Introduction}

The economy is regarded as a set of activities of irrational agents in complex networks.
However, many traditional studies in economics investigate the activities of
rational and representative agents in simple networks, i.e., regular networks and
random networks. To overcome the limitations of such an unrealistic situation, the viewpoint of
irrational agents has emerged.
However, simple networks have been adopted in economics and many studies of agent simulation.
Recently, the study of complex networks has revealed the true structure of
real-world networks \cite{barabasi2002}\cite{dm2003}.
Gross networks such as interindustry relations have been considered
in the macro economy, but the detailed structures of networks constructed
from entities in the micro economy have not been clearly elucidated \cite{sfa2003}\cite{sfa2004}\cite{sfa2005}\cite{gbcsc2003}.

By common practice, if we intend to discuss networks, we must define
their nodes and edges. Edges represent the relationships between nodes.
In this study, we consider companies as nodes.
To define the relationships between companies, we use three viewpoints: ownership,
governance, and activity.
These three viewpoints have relations with each other.
Ownership is characterized by the shareholding of companies,
which is the subject of this article.
Governance is characterized by the interlocking of directors,
and it is frequently represented by a bipartite graph constructed from corporate
boards and directors.
The activity networks are characterized by many relationships: trade, collaboration, etc.

In this article we consider Japanese shareholding network
at the end of March 2002 (see Ref.~\cite{gbcsc2003} for shareholding
networks in MIB, NYSE, and NASDAQ).
We use data published by TOYO KEIZAI INC. 
This data source provides lists of shareholders for 2,765 companies listed on
the stock market or the over-the-counter market.
The lengths of the shareholder lists vary with the companies.
The most comprehensive lists contain information on the top 30 shareholders.
Types of shareholders include listed companies,
non-listed financial institutions (commercial banks, trust banks,
and insurance companies), officers, and other individuals.
In this article, we only consider the shareholding network constructed from
2,303 companies listed on the stock market
and 53 non-listed financial institutions.
Accordingly, the size of this network is $N=2,356$, and the total number of
edges is $L=22,435$.
  
This paper is organized as follows. In the next section we consider the degree distribution
for both incoming edges and outgoing ones and show that the outgoing degree
distribution follows a power law function with an exponential cutoff.
In the following section, we discuss correlations between the degree and the company's age,
profit, and total assets.
This is because we assume that the dynamical change and growth of
business networks can be explained by the company's growth.
The last section is devoted to a summary and discussion.

%%%%%%%%%%% Derected network
\section{Degree distribution in the directed shareholding network}
If we draw arrows from shareholders to stock corporations, we can represent a
shareholding network as a directed graph.
If we count the number of incoming edges and that of outgoing edges for each node,
we can obtain the degree distribution for incoming degree, $k_\textrm{\footnotesize in}$,
and that of outgoing degree, $k_\textrm{\footnotesize out}$.
However, as explained above, the lengths of the shareholder lists vary with the
companies, and the most comprehensive lists contain the top 30 shareholders.
Therefore, the incoming degree has an upper bound,
$k_\textrm{\footnotesize in}\leq 30$, while the outgoing degree has no bound.

\begin{figure}[!t]
\centerline{
\begin{minipage}{1.\linewidth}
\includegraphics[width=\linewidth]{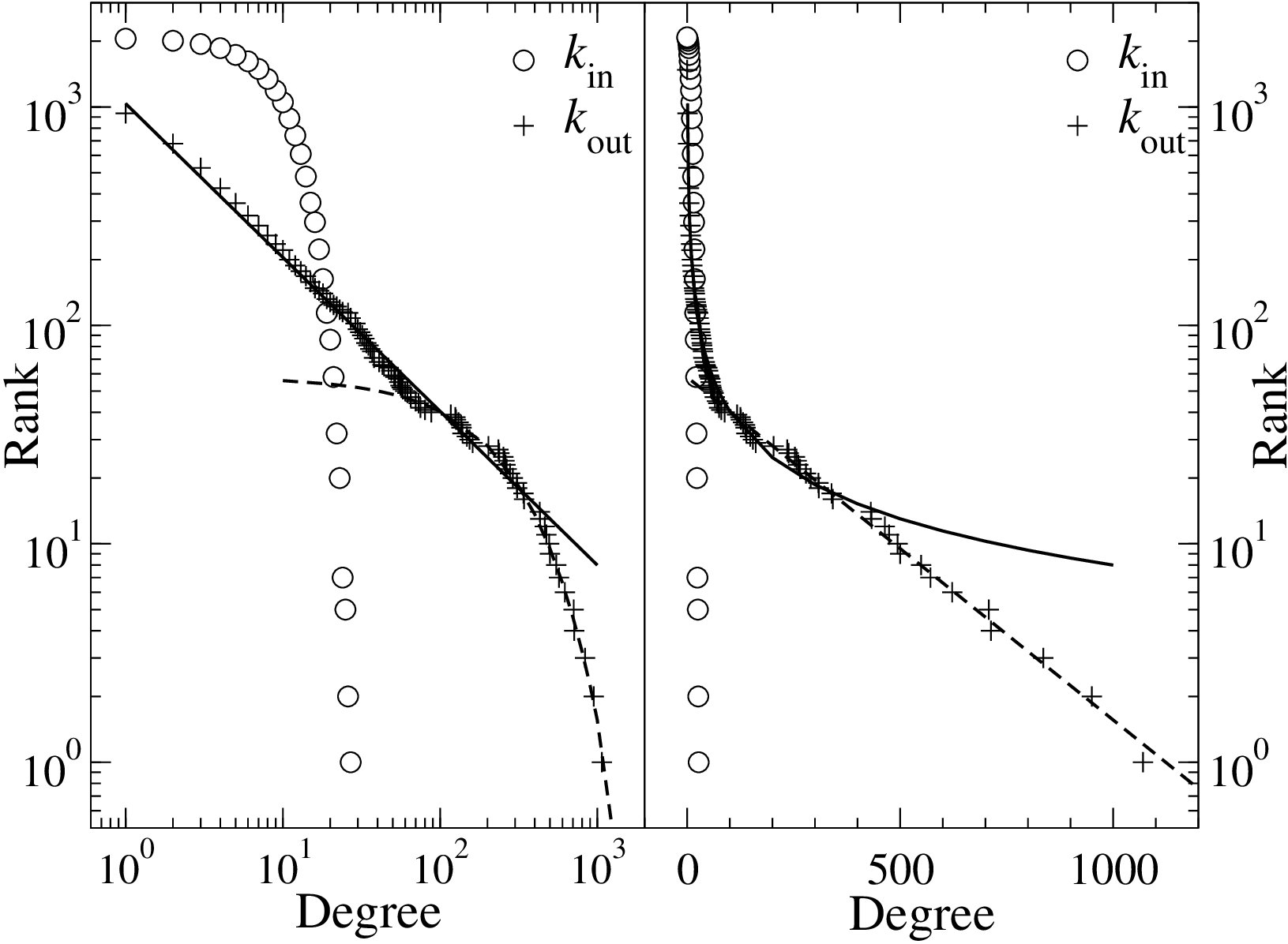}
\end{minipage}
}
\caption{Log-log plot (left) and semi-log plot (right) of the degree distribution .
The open circles represent the distribution of the incoming degree, and 
the plus symbols represent that of the outgoing degree.
The solid lines correspond to the fitting by the power law function with the exponent
$\gamma=1.7$, and the dashed lines correspond to the fitting by the exponential function.
}
\label{fig:degree}
\end{figure}

The log-log plot of the degree distribution is shown in the left
panel of Fig.~\ref{fig:degree}, and the semi-log plot is shown in the right
panel of Fig.~\ref{fig:degree}.
In this figure, the horizontal axis corresponds to the degree and
the vertical axis corresponds to the rank.
The open circles represent the distribution of the incoming degree, and 
the plus symbols represent that of the outgoing degree.
As mentioned above, we can find the upper bound for the incoming degree.
In the discussion below, therefore, we consider only the distribution of the outgoing degree.

The solid lines correspond to the fitting by the power law function 
$p(k_\textrm{\footnotesize out})\propto k_\textrm{\footnotesize out}^{-\gamma}$ with the
exponent $\gamma=1.7$,
and the dashed lines correspond to the fitting by the exponential function.
We can see that the outgoing degree distribution follows the power law distribution
with an exponential cutoff.
The exponential part of the distribution is constructed from 40 nodes, which
are 38 financial institutions and 2 trading firms.
On the other hand, the power law part of the distribution is constructed
from 2,316 nodes, and 96\% of them represent non-financial institutions.

The above results suggest that different mechanisms work in each
range of the distribution. We assume that the dynamics in the range of
the exponential distribution is essentially explained by the model proposed in Ref.~\cite{asbs2000}.
This model is based on a so-called BA model \cite{ba1999}, i.e.,
the preferential attachment in growing networks
that explains the power law distribution of the degree.
Here, we assume that the outgoing degree distribution in the tail part is explained based on
the BA model,
since financial institutions must invest money as shares. Therefore, financial institutions
actively obtain shares of companies newly listed on
the stock market or the over-the-counter market.
This mechanism is almost the same as preferential attachment.
However, we must modify the BA model in order for it to explain the exponential distribution of
the outgoing degree.
If we consider an extended BA model with aging of the nodes, then we can obtain
the exponential degree distribution \cite{asbs2000}. However, as explained in the next section,
this is not the case.
On the other hand, if we consider an extended BA model that includes the cost of adding edges to the nodes
or the limited capacity of a node, then we can obtain
the exponential degree distribution \cite{asbs2000}.
As explained in the next section, this is actually the case.

We consider cliques in networks to be important characteristics for constructing a model
that can explain the power law distribution of the outgoing degree.
Cliques in networks are quantified by a clustering coefficient \cite{ws1998}, which is
defined in the case of undirected networks.
Supposing that a node $i$ has $k_i$ edges, then at most $k_i(k_i-1)/2$ edges
can exist between them. The clustering coefficient of node $i$, $C_i$, is the fraction of
these allowable edges that actually exist for $e_i$, i.e.,
$C_i=2e_i/k_i(k_i-1)$. This has been calculated for an undirected shareholding network
by using the same data used in this article \cite{sfa2005}, and it has been shown that the
clustering coefficient follows
the power law distribution, $C(k)\propto k^{-\alpha}$, with the exponent
$\alpha=1.1$.
Such a scaling property of the distribution of clustering coefficients has also been observed in
biological networks, and this has motivated the concept of hierarchical
networks \cite{rsmob2002}\cite{rb2003}. 
We believe that the dynamics in the power law distribution of the outgoing degree
is essentially explained by extending the model proposed in Ref.~\cite{rb2003}.

The clustering coefficient is approximately equal to the probability of
finding triangles in the network. The triangle forms the minimum loop.
Therefore, if node $i$ has a small value of $C_i$, then the probability of finding loops
around this node is low.
Consequently, this scaling property of the clustering coefficient suggests
that the network is locally tree-like.

%%%%%%%%%%% Degree v.s. Company's growth
\section{Correlation between outgoing degree and company's growth}
We consider the correlations between the outgoing degree and 
the company's age, profit, and total assets.
We believe that knowing the characteristics of nodes is useful for constructing models explaining
the dynamics of networks.
In many complex networks, it is difficult to quantitatively characterize the nature of nodes.
However, in the case of economic networks, especially networks constructed from companies,
we can obtain the nature of nodes quantitatively based on balance sheets,
and income statements, for example.
We consider this a remarkable characteristic of business networks,
and it allows us to understand business networks in terms of the company's growth.
We believe that the dynamics of business networks must be explained
by the theory of company growth.

%%%%%%%%%%% Age
\subsection{Outgoing degree and company's age}
As mentioned in the previous section, we consider the correlation between
the outgoing degree and the company's age in order to clarify whether the BA model with
aging of the nodes is applicable to a particular case. In the following, we measure the company's
age in units of months.

\begin{figure}[!t]
\centerline{
\begin{minipage}{1.\linewidth}
\includegraphics[width=\linewidth]{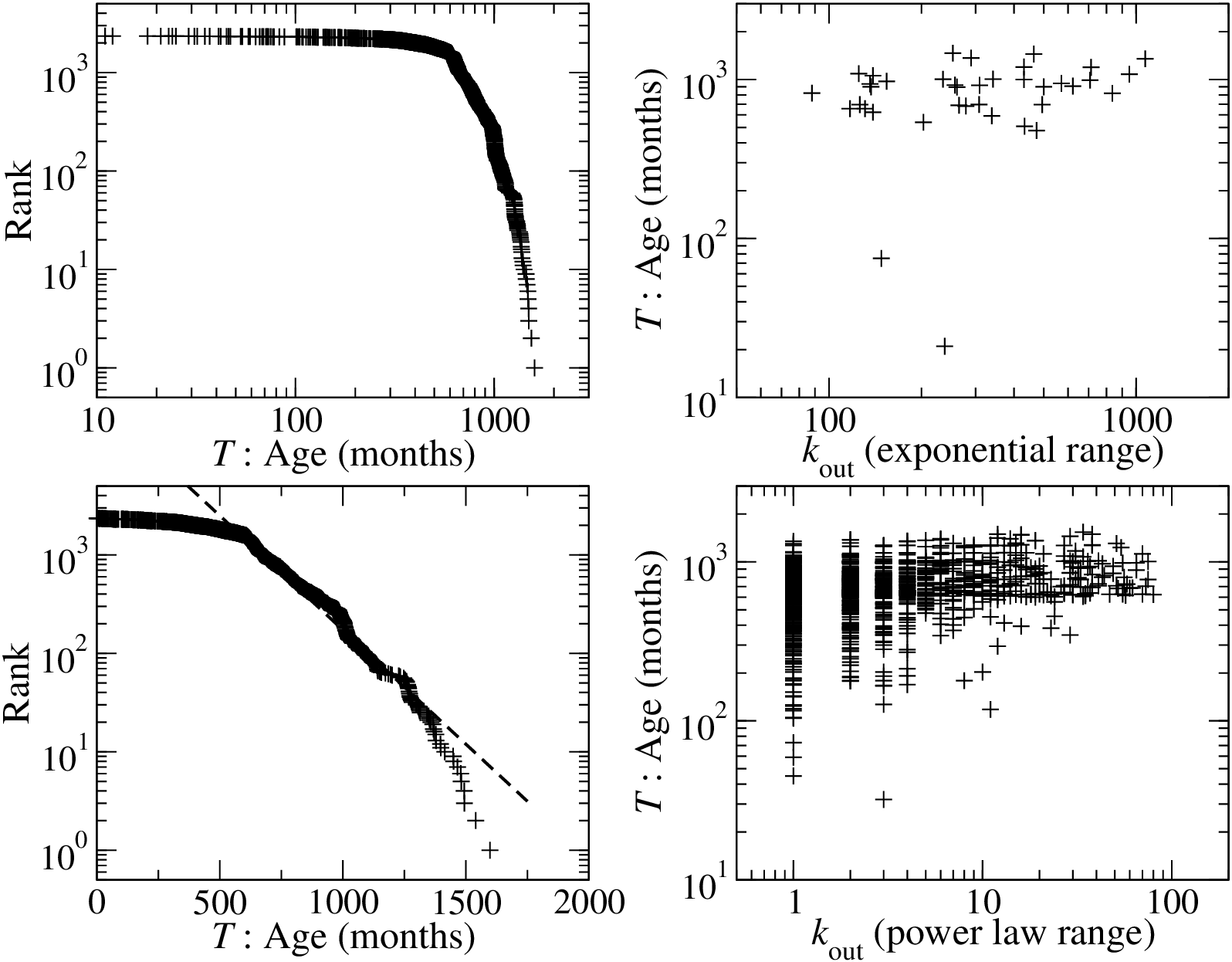}
\end{minipage}
}
\caption{Log-log plot (upper left) and a semi-log plot (lower left) of the distribution of the
company's age.
The dashed line corresponds to the fitting by the exponential function.
Log-log plots of the correlation between the company's age and the outgoing degree
that follows the exponential distribution (upper right),
and the one that follows the power law distribution (lower right).
}
\label{fig:age}
\end{figure}

The log-log plot of the distribution of the company's age is shown in the upper left panel
of Fig.~\ref{fig:age}. In this figure, the horizontal axis is the age in the unit of months,
and the vertical axis is the rank. This figure shows that the distribution 
does not fit a simple function such as a power law function.

The semi-log plot of the distribution of the company's age is shown in the lower left panel
of Fig.~\ref{fig:age}. In this figure, the meaning of the axes is the same as in the upper figure.
The dashed line corresponds to the fitting by the
exponential function.
This means that the distribution of the company's age approximately follows
the exponential distribution.
It is expected that the age of a company has a relation with its lifetime, and
the lifetime of bankrupted companies follows an exponential
distribution \cite{fujiwara2004}\cite{fagsg2004}.

The upper right panel of Fig.~\ref{fig:age} shows a log-log plot of the correlation
between the company's age in months and the outgoing degree that
follows the exponential distribution. We can observe that these
two quantities have no correlation. To quantify this observation, we calculated Spearman's
rank correlation $S$ and Kendall's rank correlation $K$ and obtained $S=0.27$ and $K=0.17$,
respectively. These results mean that there is no correlation between the
company's age and the outgoing degree that follows the exponential distribution.
This suggests that the BA model with aging of the nodes is not applicable
to this case.

The lower right panel of Fig.~\ref{fig:age} shows a log-log plot of the correlation
between the company's age in months and the outgoing degree that
follows the power law distribution. We can also observe that these two quantities have
no correlation.
In this case, $S=0.32$ and $K=0.24$.
These results mean that there is no correlation between the company's age and the
outgoing degree that follows the power law distribution.
Needless to say, the BA model with aging of the nodes is not applicable
to this case.

A comparison of these two figures clarifies the independence between
the outgoing degree and the company's age. There are some companies with a few outgoing
edges in the range where the lifetime is longer than $10^3$ months.

%%%%%%%%%%% Profit 
\subsection{Outgoing degree and company profit}
We consider the correlation between the outgoing degree and the company's profit.
Here the company's profit is the amount of money that the company gains when it
is paid more for something than it cost to make, get or do it.
Hence the profit includes information on the flow in the network.
Here, we consider the company's profit as an amount of money stored
in the period from the beginning of April 2001 to the end of March 2002.

\begin{figure}[!t]
\centerline{
\begin{minipage}{1.\linewidth}
\includegraphics[width=\linewidth]{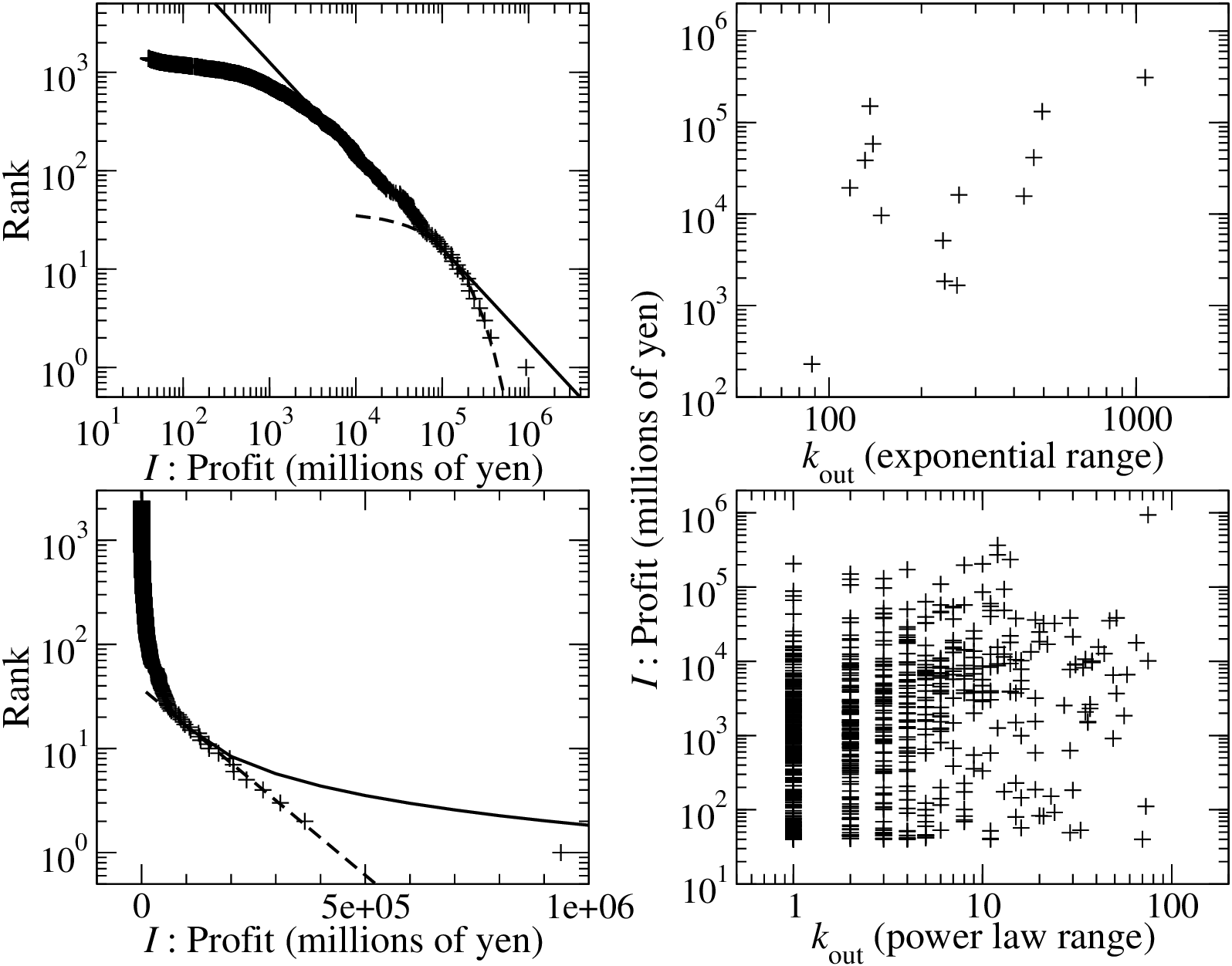}
\end{minipage}
}
\caption{Log-log plot (upper left) and a semi-log plot (lower left) of the distribution of
the company's profit.
The solid lines correspond to the fitting by the power law distribution with the
exponent $a=2$, and the dashed lines correspond to the fitting by the exponential function.
Log-log plots of the correlation between the company's profit and the outgoing degree
that follows the exponential distribution (upper right)
and the one that follows the power law distribution (lower right).
}
\label{fig:profit}
\end{figure}

The log-log plot of the distribution of the company's profit is shown in the
upper left panel of Fig.~\ref{fig:profit}. In this figure, the horizontal axis is profit $I$ in
units of million yen, and the vertical axis is the rank. The solid line corresponds to the fitting by the
power law function. This is represented by the probability density function (pdf), $p(I)$ as
$p(I)\propto I^{-a}$ with the exponent $a=2$.
This shows that the distribution in the profit in the middle range follows the power law
distribution. The dashed line corresponds to the fitting by the exponential function.
The exponential distribution in the high-profit range is better clarified by the semi-log plot.

The semi-log plot of the distribution of the company's profit is shown in the lower left panel
of Fig.~\ref{fig:profit}. The meanings of axes, solid line, and dashed line
are the same as those in the upper panel. This figure shows that the profit in the tail part
follows an exponential distribution.
However, the number of companies within this range is small.

The upper right panel of Fig.~\ref{fig:profit} shows a log-log plot of the correlation
between the company's profit in the unit of million yen
and the outgoing degree that follows the exponential distribution.
As mentioned previously, the exponential part of the outgoing degree distribution
contains 40 nodes, which are 38 financial institutions and 2 trading firms.
Almost all of these financial institutions are not listed on the stock market, and we could not
obtain their profit data. Needless to say, there is also a possibility that the profit of these
financial institutions was negative.
Therefore, the total number of data point is less than 40.
We can see that these two quantities have no or only weak correlation.
Spearman's rank correlation and Kendall's rank correlation are $S=0.31$ and $K=0.23$,
respectively. These results mean that there is no or only weak correlation between the
company's profit and the outgoing degree that follows the exponential distribution.

The lower right panel of Fig.~\ref{fig:profit} shows a log-log plot of the correlation
between the company's profit and the outgoing degree that follows the power law
distribution. We can see that these two quantities also have no or only weak correlation.
In this case, $S=0.32$ and $K=0.24$.
These results mean that there is no or only weak correlation between the company's profit and the
outgoing degree that follows the power law distribution.

Here, we consider the profits of the companies comprising the network.
Therefore, the distribution of profit, i.e., the upper left panel of Fig.~\ref{fig:profit},
is not so impressive. However, the profit of Japanese companies has remarkable
properties: (i) It is confirmed that in the period after 1970
the high-profit range followed the power law $p(I)\propto I^{-2}$, i.e., Pareto-Zipf's law,
and was stable \cite{mktt2002}.
(ii) It is confirmed for the years 2000 and 2001 that detailed balance was maintained for
the top 70,000 companies in the high-profit range \cite{asf2003}.
(iii) It is also confirmed for the years 2000 and 2001 that Gibrat's law was maintained for
the top 70,000 companies in the high-profit range \cite{asf2003}.
Here, Gibrat's law means that the growth rate of profit is independent of
the profit in the initial year. (iv) These characteristics are related to each other \cite{asf2003}.

%%%%%%%%%%% Asset
\subsection{Outgoing degree and company's total assets}
We next considered the correlation between the outgoing degree and the company's total assets.
The company's total assets are naively regarded as the amount of profit.
A portion of the company's total assets is invested in shares.
Accordingly, it is expected that the company's total assets have a relation with
the cost of adding edges to the nodes or with the limited capacity of a node.

\begin{figure}[!t]
\centerline{
\begin{minipage}{1.\linewidth}
\includegraphics[width=\linewidth]{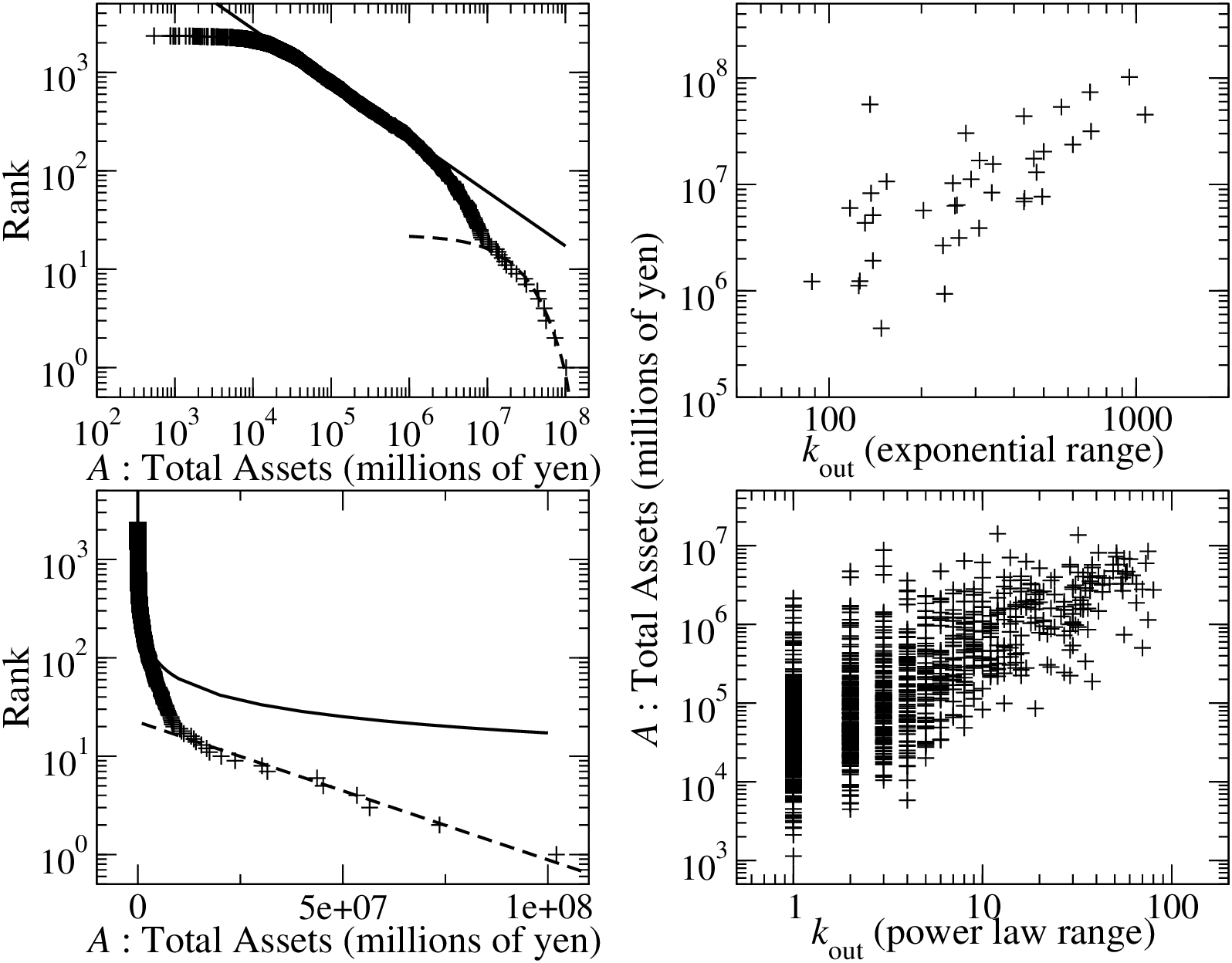}
\end{minipage}
}
\caption{Log-log plot (upper left) and
a semi-log plot (lower left) of the distribution of the company's total assets .
The solid lines correspond to the fitting by the power law distribution with
the exponent 1.6, and the dashed lines correspond to the fitting by the exponential function.
Log-log plots of the correlation between the company's total assets and the outgoing degree
that follows the exponential distribution (upper right),
and the one that follows the power law distribution (lower right).
}
\label{fig:asset}
\end{figure}

The log-log plot of the distribution of the company's total assets at the end of March 2002
is shown in the upper left panel of Fig.~\ref{fig:asset}. In this figure, the horizontal axis is the
total assets $A$ in the units of million yen, and the vertical axis is the rank. The solid line corresponds
to the fitting by the power law function $p(A)\propto A^{-b}$ with the exponent $b=1.6$.
This figure shows that the level of total assets in the middle range follows the power law distribution.
The dashed line corresponds to the fitting by the exponential function.

The semi-log plot of the distribution of the company's total assets is shown in the lower left panel
of Fig.~\ref{fig:asset}. The meanings of axes, solid line, and dashed line
are the same as in the upper panel. This figure shows that the level of the total assets in the tail
part of the distribution follows the exponential distribution.
However, the change from the power law distribution to the
exponential distribution is not smooth.

The upper right panel of Fig.~\ref{fig:asset} shows a log-log plot of the correlation
between the company's total assets at the end of March 2002 in units of million yen
and the outgoing degree that follows the exponential distribution.
Contrary to the case of the company's profit, we can obtain data on the company's
total assets for the non-listed financial institutions.
This figure shows that these two quantities strongly correlate.
Spearman's rank correlation and Kendall's rank correlation are $S=0.71$ and $K=0.55$,
respectively. These results mean that there is strong correlation between the company's
total assets and the outgoing degree that follows the exponential distribution.
This suggests that the BA model incorporating the cost of adding edges to the nodes
or the limited capacity of a node is applicable to this case.

The lower right panel of Fig.~\ref{fig:asset} shows a log-log plot of the correlation
between the company's total assets and the outgoing degree that follows the power law
distribution. We can see that these two quantities also have strong correlation.
In this case, $S=0.65$ and $K=0.51$.
These results mean that there is strong correlation between the company's total assets and the
outgoing degree that follows the power law distribution.

This suggests that the level of the company's total assets has strong correlation with the outgoing
degree. In particular, it is assumed that the outgoing degree distribution in the tail part
is explained by the BA model incorporating the cost of adding edges to the nodes
or the limited capacity of a node.
However to confirm this assumption, we must study
the accumulation process of the company's total assets.

\section{Summary}
In this paper we considered the Japanese shareholding network existing at the end of March 2002.
Although we could not clarify the detailed characteristics of the incoming
degree distribution, we could obtain useful information on the outgoing degree distribution:
(i) The outgoing degree distribution follows the power law with an exponential cutoff.
(ii) The important factor in the growth of the business network is not the company's age
but its total assets.
This means that old companies do not necessarily have large total assets.
(iii) It is expected that the dynamics of the tail part of the outgoing degree distribution, i.e. the
exponential distribution range, can be explained by the BA model incorporating the cost of adding edges to
the nodes or the limited capacity of a node.
However, the mechanism for the power law part of
the outgoing degree distribution is still not known.
We believe that the dynamical change and growth of
business networks must be explained by the company's growth.
Consequently, knowing the dynamics of the company's growth is a key concept in considering
the growth of economic networks \cite{fgags2004}\cite{afs2004}.

We would like to conclude with two observations. 
The first concerns degree correlation.
In this article, we considered the shareholding network as a directed graph.
However, if we ignore the direction of edge, we can calculate
many quantities to obtain the characteristics of networks.
For example, the nearest neighbors' average degree of nodes with degree $k$,
$\langle k_{\rm nn}\rangle(k)$, is an important quantity \cite{pvv2001}.
This was calculated by Ref.~\cite{sfa2005} for the Japanese shareholding networks at the
end of March 2002, in which $\langle k_{\rm nn}\rangle(k) \propto k^{-\nu}$
was obtained with the exponent $\nu=0.8$.
This means that hubs are not directly connected to each other
in this network, i.e., it's a degree non-assortative network.

The second observation concerns the spectrum of the graph.
It has recently been shown through an effective medium approximation that
the probability density function of the eigenvalue for the adjacency matrix, $\rho(\lambda)$,
is asymptotically represented by that of the degree distribution $p(k)$, i.e.,
$\rho(\lambda)\simeq 2\left|\lambda\right| p(\lambda^2)$,
if the network has a local tree-like structure \cite{dgmc2003}.
As mentioned previously, the shareholding network has this characteristic.
Therefore, if $p(k)$ asymptotically follows the power law distribution,
$\rho(\lambda)$ also asymptotically follows the power law distribution.
In addition, we can derive the scaling relation $\delta=2\gamma-1$.
At the end of March 2002, the tail part of the eigenvalue for the adjacency matrix
of the undirected Japanese shareholding network followed
$\rho(\lambda)\propto |\lambda|^{-\delta}$ with the exponent $\delta=2.6$.
On the other hand, the degree distribution in this case is
$p(k)\propto k^{-\gamma}$ with the exponent $\gamma=1.8$.
Therefore, the scaling relation is guaranteed \cite{sfa2005}.
%%%%%%%%%%%%%%%%%%%%%%%%%%%%%%%%%%%%%%%%%%%%%%%%
%% BACKMATTER
%%%%%%%%%%%%%%%%%%%%%%%%%%%%%%%%%%%%%%%%%%%%%%%%

\begin{theacknowledgments}
This work is supported in part by the National Institute of Information
and Communications Technology.
We are also supported in part by a Grant-in-Aid for Scientific
Research (\#15201038) from the Ministry of Education, Culture,
Sports, Science and Technology. We would like to thank Dr. Katsunori Shimohara for
his encouragement.
\end{theacknowledgments}



\bibliographystyle{aipproc}   % if natbib is available

\bibliography{sample}

\begin{thebibliography}{99.}
\bibitem{barabasi2002}
A.-L. Barab\'{a}si,
\emph{Linked: The New Science of Networks},
Perseus Press, Cambridge, MA, 2002.

\bibitem{dm2003}
S.~N.~Dorogovtsev, and J.~F.~F.~Mendes,
\emph{Evolution of Networks: From Biological Nets to the Internet and WWW},
Oxford University Press, Oxford, 2003.

\bibitem{sfa2003}
W.~Souma, Y.~Fujiwara, and H.~Aoyama,
\emph{Physica A}, \textbf{324}, 396--401 (2003).

\bibitem{sfa2004}
W.~Souma, Y.~Fujiwara, and H.~Aoyama,
\emph{Physica A}, \textbf{344}, 73--76 (2004).

\bibitem{sfa2005}
W.~Souma, Y.~Fujiwara, and H.~Aoyama,
``Heterogeneous economic networks,'' in \emph{Economics and Heterogeneous
Interacting Agents}, co-edited by A.~Namatame, et~al.,
Springer-Verlag, Tokyo, 2005, to be published. arXiv:physics/0502005.

\bibitem{gbcsc2003}
D.~Garlaschelli, et~al.,
to be published in \emph{Physica A}. arXiv:cond-mat/0310503.

\bibitem{asbs2000}
L.~A.~N.~Amaral, et~al.,
\emph{PNAS}, \textbf{97}, 11149--11152 (2000).

\bibitem{ba1999}
A.-L.~Barab\'{a}si, and R.~Albert,
\emph{Science}, \textbf{286}, 509--512 (1999).

\bibitem{ws1998}
D.~J.~Watts, and S.~H.~Strogatz,
\emph{Nature}, \textbf{393}, 440--442 (1998).

\bibitem{rsmob2002}
E.~Revasz, et~al.,
\emph{Science}, \textbf{297}, 1551--1555 (2002).

\bibitem{rb2003}
E.~Revasz, and A.-L.~Barab\'{a}si,
\emph{Phys. Rev. E}, \textbf{67}, 026112 (2003).

\bibitem{fujiwara2004}
Y.~Fujiwara,
\emph{Physica A}, \textbf{337}, 219--230 (2004).

\bibitem{fagsg2004}
Y.~Fujiwara, et~al.,
\emph{Physica A}, \textbf{344}, 112--116 (2004).

\bibitem{mktt2002}
T.~Mizuno, et~al.,
``Statistical laws in the income of Japanese Companies,'' in
\emph{Empirical Science of Financial Fluctuations: The Advent of Econophysics},
edited by H.~Takayasu, Springer-Verlag, Tokyo, 2002, pp. 321--330.

\bibitem{asf2003}
H.~Aoyama, W.~Souma, and Y.~Fujiwara,
\emph{Physica A}, \textbf{324}, 352--358 (2003).

\bibitem{fgags2004}
Y.~Fujiwara, et~al.,
\emph{Physica A}, \textbf{335}, 197--216 (2003).

\bibitem{afs2004}
H.~Aoyama, Y.~Fujiwara, and W.~Souma,
\emph{Physica A}, \textbf{344}, 117--121 (2004).

\bibitem{pvv2001}
R.~Pastor~Satorras, A.~V\'{a}zquez, and A.~Vespignani,
\emph{Phys. Rev. Lett.}, \textbf{87}, 258701 (2001).

\bibitem{dgmc2003}
S.~N.~Dorogovtsev, et~al.,
\emph{Phys. Rev. E}, \textbf{68}, 046109 (2003).

\end{thebibliography}

\IfFileExists{\jobname.bbl}{}
 {\typeout{}
  \typeout{******************************************}
  \typeout{** Please run "bibtex \jobname" to optain}
  \typeout{** the bibliography and then re-run LaTeX}
  \typeout{** twice to fix the references!}
  \typeout{******************************************}
  \typeout{}
 }

\end{document}